# Vector cnoidal and solitary plasmon polariton waves in a planar waveguide

## IGOR V. DZEDOLIK

*V. I. Vernadsky Crimean Federal University, 4 Vernadsky Avenue, Simferopol, 295007, Russian Federation*
*Corresponding author: igor.dzedolik@cfuv.ru*

**The paper considers the dynamics of nonlinear surface plasmon polariton waves in a planar plasmon waveguide, which is a heterostructure of non-magnetic metallic and dielectric layers. The obtained in the work nonlinear equations and their analytical solutions describe the vector cnoidal and solitary plasmon polariton waves excited by single electromagnetic pulse at the waveguide. Nonlinear plasmon polariton waves arise under the influence of the Kerr nonlinearity of metal and the saturation of nonlinearity at the heterostructure. The period and profile of envelope of the excited nonlinear surface plasmon polariton wave vary depending on the conditions of excitation and the power of exciting electromagnetic pulse.**

## 1. INTRODUCTION

Nanoscale plasmonic systems attract nowadays the special attention of researchers in connection with unique properties of the plasmonic units for the purpose of their applications to the subwavelength integrated circuits operating at optical frequencies [1,2]. Plasmonic nanodevices based on the linear and nonlinear effects at the optical frequencies are able to reproduce a large number of functions of semiconductor electronic devices operating at the gigahertz frequencies. In this regard, the use of surface plasmon polariton periodically and solitary waves is becoming more and more popular due to their universal properties of signal transmission and operation at the optical frequencies [3-5].

The components of plasmonic circuits, particularly, the logic gates using the nonlinear properties [6], plasmonic modulators [7], plasmon field effect transistors [8] are now designed based on the dielectric-metal-dielectric and metal-dielectric-metal heterostructures. The realization of these plasmonic units requires the transmission and processing of the optical frequency signals with the low and high intensity between the plasmonic components inside a microchip. Under the influence of high intensity optical signals, the nonlinear properties of media change the parameters of transmitted signal [3-7,9-13]. The nonlinear response of noble metals applying in the plasmonic circuits attracts the special attention of researches to investigate them theoretically and experimentally [12]. Because of nonlinear response, the self-modulation of the surface plasmon polariton wave occurs [14-19]. Depending on the surface plasmon polaritons (SPPs) intensity, the properties of plasmon signals are changed under the influence of nonlinearity of the dielectric-metal interface, particularly, the wavelength of the transmitted plasmon signals can vary [17,18].

The optical signals are transmitted inside the microchip along the metal waveguides with subwavelength sizes by the generation therein the nonlinear plasmon polaritons pulses [20-26]. The conduction currents in the volume and at the boundaries of the waveguide do not arise. This phenomenon makes it possible to generate SPPs at optical frequencies, which allows significantly expand the signal transmission band in plasmonic devices.

SPPs arise as the mixture of photons, phonons and plasmons in both dielectric and metallic media. These SPPs are "attached" to the interface of the media, and they are not radiated from the smooth interface [1]. The wavevector of SPPs is several times greater than the wavevector of the exciting optical wave in air or in optical fibers of the same frequency. The short SPP wavelength is one more advantage of plasmonics for the production of nanodevices.

The excitation of planar plasmon waveguides in microchip is usually realized either with the help of tapered metallized optical fiber tips at telecommunication wavelengths or with the use of spaser-type optical radiation generators built into a chip, in particular, in the visible range [6]. In these cases the nonlinear response of the waveguide media depends on the carrier wavelength (frequency) of the exciting electromagnetic signal. Powerful electromagnetic pulse generates nonlinear plasmon polariton waves called the cnoidal waves and solitary waves as solitons at the interface between conducting and dielectric media [20-26]. The parameters and dynamics of plasmon cnoidal waves and solitons depend on the intensity of the exciting electromagnetic pulses, as well as on the properties of the media and on the geometry of the waveguide in which plasmon polaritons are propagating.

In the work, there is the theoretically analysis of the varying of SPP dynamics depending on the nonlinear properties of the planar dielectric-metal-dielectric waveguide in the differ regimes of SPP pulse propagation. The terms "strong nonlinear response" and "weak nonlinear response" were introduced to define the dependence of the nonlinear permittivity of media on the power of SPP pulse. SPP properties of plasmonic waveguide are described by the system of nonlinear equations with the saturation of the Kerr-type nonlinearity.

The power short electromagnetic pulse generates the vector nonlinear SPP wave due to the Kerr-type nonlinearity with saturation at "strong nonlinear response" of medium. Also, the SPP pulse is propagating along the waveguide in the form of vector cnoidal wave or solitary wave due to the Kerr-type nonlinearity of the media at "weak nonlinear response". Thus, the solitary short electromagnetic pulse can excite either the cnoidal wave or the solitary wave in the plasmon waveguide depending on the ratio of parameters of the exciting pulse and the waveguide. This case differs from the theoretically investigated case of two-pulse "pump-probe" excitation of SPPs in the plasmon waveguide [23,24].

This theoretical analysis will allow one to design and implement the subwavelength components of plasmonic nanocircuits operating at optical frequencies.

## 2. NONLINEAR EVANESCENT WAVES OF SURFACE PLASMON POLARITONS

The SPP wave in a plasmon waveguide can be excited by a short electromagnetic pulse at the cases of "strong nonlinear response" and "weak nonlinear response". Let us consider a nonlinear evanescent SPP wave that is propagating along the $z$-axis at the interface of media as metal layer with the permittivity $\varepsilon_M$ embedded into the dielectric medium with the permittivity $\varepsilon_D$ which represent a metal non-magnetic planar plasmon waveguide. Assume that the amplitude of the wave of nonlinear surface plasmon-polaritons (NSPPs) decreases exponentially with the distance from the interface along the $x$-axis that is normal to the boundary as $\sim \exp(-\alpha_D x)$ in the positive direction $x > 0$ (in the dielectric), and in the negative direction $x < 0$ (in the metal) as $\sim \exp(\alpha_M x)$, and $\text{Re}\alpha_D > 0$, $\text{Re}\alpha_M > 0$ (Fig. 1).

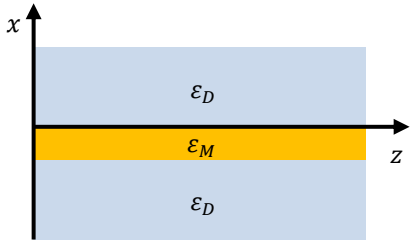

Fig. 1. Surface of a plasmon waveguide (side view).

Similar NSPP waves can be excited inside a planar plasmon waveguide, which is a metal-dielectric-metal heterostructure [1,2].

The system of equations for the components of electric and magnetic vector of the NSPP wave excited in the TM mode $E_x, E_z, B_y \sim \exp(\alpha x)$ with plane front at the dielectric (D) or in the metal (M) has the form (APPENDIX A)

$$\frac{1}{c}\frac{\partial(\varepsilon E_x)}{\partial t} + \frac{\partial B_y}{\partial z} = 0, \frac{1}{c}\frac{\partial(\varepsilon E_z)}{\partial t} - \alpha B_y = 0, \frac{\partial E_x}{\partial z} - \alpha E_z + \frac{1}{c}\frac{\partial B_y}{\partial t} = 0, \quad (1)$$

where $\varepsilon(\tau) = \varepsilon_L + \chi(E_x^* E_x + E_z^* E_z)$, $\varepsilon_L$ is the dielectric constant, $L = (D, M)$, $\chi = (4\pi\chi_D, 4\pi\chi_M)$ is the Kerr susceptibility, $\alpha = (-\alpha_D, \alpha_M) = const$ is the transverse damping coefficient of NSPP in the medium. The nonlinear susceptibility of the medium $\chi$ depends on the frequency of the exciting electromagnetic field [20].

## 3. VECTOR CNOIDAL AND SOLITARY WAVES

For the analytical analysis of the properties of the NSPP wave excited by the short electromagnetic pulse, let us consider the case of "weak nonlinear response". The self-modulation of NSPP wave occurs in Kerr media under the influence of nonlinearity at sufficient power of the exciting electromagnetic signal, that leads to the generation of cnoidal waves (nonlinear periodic waves) or solitary waves (solitons) [5]. The period and profile of the cnoidal and solitary waves depend on the energy density of the exciting electromagnetic pulse.

When the plasmon waveguide is excited by an electromagnetic pulse, it is convenient to pass to the moving frame of reference with the delayed time $\tau = t - z/v$, where $v = const$. The system of equations for the components of the electric vector of NSPP of the TM-mode in the moving frame of reference in the dielectric or in the metal takes the form (APPENDIX A)

$$\frac{dE_x}{d\tau} + \frac{1}{\varepsilon - n^2}\frac{d\varepsilon}{d\tau}E_x - \frac{\alpha c n}{\varepsilon - n^2}E_z = 0, \quad \frac{dE_z}{d\tau} + \frac{1}{\varepsilon}\frac{d\varepsilon}{d\tau}E_z - \frac{\alpha c}{n}E_x = 0, \quad (2)$$

where $\varepsilon(\tau) = \varepsilon_L + \chi(E_x^* E_x + E_z^* E_z)$, $n = c/v$. The denominators $\varepsilon - n^2$ and $\varepsilon$ in the system of equations (2) depend on the NSPP energy density $w \sim E^2 = E_x^* E_x + E_z^* E_z$, i.e. they describe the saturation of the medium nonlinearity. The magnetic component of TM mode is $B_y = \frac{1}{\alpha c}\frac{\partial}{\partial \tau}(\varepsilon E_z)$.

For analytical analysis of the dynamics of excited the NSPP waves, let us consider the system of nonlinear equations (2) in the case of a low intensity of the NSPP wave $\chi E^2 \ll 1$ without taking into account losses in the medium. Assuming $\left|\frac{dE_x}{d\tau}\right| \gg \left|\frac{1}{\varepsilon - n^2}\frac{d\varepsilon}{d\tau}E_x\right|$ and $\left|\frac{dE_z}{d\tau}\right| \gg \left|\frac{1}{\varepsilon}\frac{d\varepsilon}{d\tau}E_z\right|$, we obtain from the system of equations (1) the system of symmetric nonlinear equations

$$\frac{d^2 E_x}{d\tau^2} - \frac{\Omega^2}{\Delta\varepsilon + \chi E^2}E_x = 0, \quad \frac{d^2 E_z}{d\tau^2} - \frac{\Omega^2}{\Delta\varepsilon + \chi E^2}E_z = 0, \quad (3)$$

where $\varepsilon = \varepsilon_L + \chi E^2$, $E^2 = E_x^2 + E_z^2$, $\Delta\varepsilon = \varepsilon_L - n^2$, $\Omega^2 = \alpha^2 c^2$.

The analytical solution of the system of equations (3) can be found if we represent the system in the form of two equations of motion [27]

$$\frac{d^2 E_x}{d\tau^2} = -\frac{\partial U}{\partial E_x}, \quad \frac{d^2 E_z}{d\tau^2} = -\frac{\partial U}{\partial E_z}, \quad (4)$$

where $U = -\frac{\Omega^2}{2\chi}\ln(\Delta\varepsilon + \chi E^2)$ is the effective potential. The electric vector of the NSPP wave has two components

$$\boldsymbol{E} = \boldsymbol{1}_x E_x + \boldsymbol{1}_z E_z.$$

We represent the system of equations (4) in the form of vector equation

$$\frac{d^2\mathbf{E}}{d\tau^2} = -\frac{\partial U}{\partial \mathbf{E}}. \quad (5)$$

The vector $\mathbf{E}$ belongs to the plane $(x, z)$ that is normal to the interface of media. Taking into account that $\partial/\partial \mathbf{E} \to (\partial\tau/\partial \mathbf{E})d/d\tau$, we obtain the first integral of the vector equation (5)

$$\frac{1}{2}\left(\frac{d\mathbf{E}}{d\tau}\right)^2 + U(\varepsilon) = \tilde{E}, \quad (6)$$

where $\tilde{E} = const$. In a moving frame of reference, the Eq. (6) coincides with the equation following from the conservation law of energy $\tilde{E}$ for a material point with unit mass moving along a flat trajectory in the central field.

We can represent the solution of Eq. (6) in the form
$$\mathbf{E} = E(\mathbf{1}_x \cos\varphi + \mathbf{1}_z \sin\varphi),$$
where $E(\tau)$ is the module of the NSPP electric vector, $\varphi = \varphi(\tau)$ is the cyclic coordinate (phase) of the NSPP wave. Then Eq. (6) takes the form

$$\left(\frac{dE}{d\tau}\right)^2 + E^2\left(\frac{d\varphi}{d\tau}\right)^2 = 2(\tilde{E} - U). \quad (7)$$

The Eq. (7) can be represented in the form of the Hamilton equations $\frac{\partial p_E}{\partial \tau} = -\frac{\partial \tilde{H}}{\partial E}$ and $\frac{\partial p_\varphi}{\partial \tau} = \frac{\partial \tilde{H}}{\partial \varphi} = 0$ with the Hamilton function $\tilde{H} = \frac{1}{2}\left(p_E^2 + \frac{1}{E^2}p_\varphi^2\right) + U$, where $p_E = \frac{dE}{d\tau}$ and $p_\varphi = E^2\frac{d\varphi}{d\tau}$ are the generalized momenta [27]. As it follows from the second Hamilton equation, the azimuthal momentum $p_\varphi = \tilde{P} = $ const is an integral of motion. Then we rewrite Eq. (7) in the form of two equations for the variables $E$ and $\varphi$,

$$\left(\frac{dE}{d\tau}\right)^2 = 2\tilde{E} + \frac{\Omega^2}{\chi}ln(\Delta\varepsilon + \chi E^2) - \frac{1}{E^2}\tilde{P}^2, \quad (8)$$

$$\frac{d\varphi}{d\tau} = \frac{1}{E^2}\tilde{P}^2, \quad (9)$$

The Eq. (8) defines the generalized phase trajectories of the NSPP on the plane $(E, dE/d\tau)$.

The module of NSPP electric vector at the dielectric-metal interface has the form of the cnoidal wave at exciting NSPP by the electromagnetic pulse (APPENDIX B)

$$E = \left[\frac{a_3 - a_2 sn^2(\bar{\tau}, \tilde{k})}{cn^2(\bar{\tau}, \tilde{k})}\right]^{1/2}, \quad (10)$$

where $sn(\bar{\tau}, \tilde{k})$ and $cn(\bar{\tau}, \tilde{k})$ are the Jacobi elliptic sine and cosine, $\bar{\tau} = \frac{\sqrt{\chi(a_1-a_3)}}{\sqrt{2}\Delta\varepsilon}\Omega\tau$, $v = \frac{c}{[\varepsilon_L - \chi(a_1+a_2+a_3)/2]^{1/2}}$ is the velocity of NSPP wave, $\tilde{k} = \sqrt{\frac{a_1-a_2}{a_1-a_3}}$ is the modulus of an elliptic integral of the first kind, where $a_1 > a_2 > a_3 > E^2$.

In case $a_1 > a_2 = a_3 > E^2$, we obtain the solitary wave excited by the electromagnetic pulse at the dielectric-metal interface

$$E = [a_1 sch^2(\bar{\tau}) + a_2 tanh^2(\bar{\tau})]^{1/2}. \quad (11)$$

The cnoidal wave Eq. (10) and the solitary wave Eq. (11) of NSPP are represented in Fig. 2.

Thus, when the planar plasmonic waveguide is excited by a pulsed electromagnetic signal, the envelope of the NSPP pulse takes the form either of cnoidal wave Eq. (10) or solitary wave Eq. (11). The solitary wave is the sum of light and dark solitons arising under self-modulation of the NSPP flow. The dynamics of the projections of NSPP electric vector $E_x$ and $E_z$ on the coordinate axis depends on the phase $\varphi$ of the plasmon polariton wave, and this phase is determined by solution of the Eq. (9) after substitution of the Eq. (10) or Eq. (11).

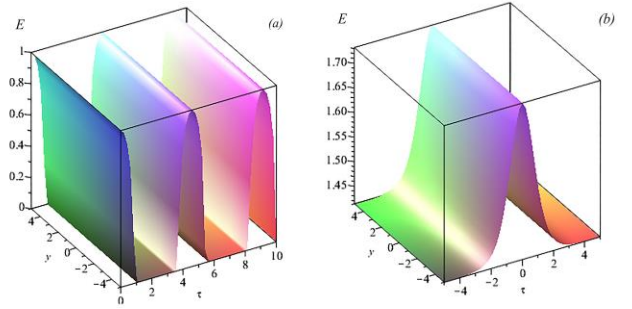

Fig. 2. Module of NSPP electric vector: (a) cnoidal wave; (b) solitary wave; arbitrary units.

The module $E$ and phase $\varphi$ of the NSPP wave depend on the constants $a_1, a_2, a_3$ (APPENDIX B). Vice versa, the constants $a_1, a_2, a_3$ depend on the integrals of motion $\tilde{E}$ and $\tilde{P}$, which are determined by the values of the parameters of the exciting electromagnetic field and media. Then, the profile and period of the NSPP wave depend on this parameters $\tilde{E} = \frac{1}{2}\left(\frac{dE(0)}{d\tau}\right)^2 + \frac{\Omega^2}{2\chi}ln[\Delta\varepsilon + \chi E^2(0)]$ that is the energy, as well as $\tilde{P} = E^2(0)\,d\varphi(0)/d\tau$ that is the azimuthal momentum at $t = 0$ and $z = 0$. When the modulus $\tilde{k}$ of the elliptic integral tends to unit, the profile of cnoidal wave is steepening and the period become longer. In this case, the NSPP pulse excited by the electromagnetic pulse propagates along the axis z of the waveguide as the soliton, and the constants are equal $a_2 = a_3 = \left(b_2 + \sqrt{3b_1 + b_2^2}\right)/3$, where $b_1 = \left[2\tilde{E} + \frac{\Omega^2}{\chi}ln(\Delta\varepsilon)\right]\frac{2(\Delta\varepsilon)^2}{\chi\Omega^2}$, $b_2 = \frac{2\Delta\varepsilon}{\chi}$ (APPENDIX B).

## 4. DISCUSSION

The cnoidal or solitary wave of NSPPs are excited in the planar plasmon waveguide by short electromagnetic pulse with sufficient power. As it is shown theoretically in Sec. 3, the cnoidal wave in the waveguide arises even when it is excited by the single electromagnetic pulse. In this case, if the power of the exciting pulse is nonsufficient SPP pulse is excited and propagated along the plasmon waveguide in the Gaussian profile. This circumstance makes it possible to determine experimentally the conditions for the excitation of nonlinear SPP waves in the plasmonic waveguide.

In order to provide the experiment on the excitation of cnoidal NSPP waves in the planar plasmon waveguide one can use the heterostructure of two diffraction gratings that are milled in a metal (gold) layer deposited on a glass substrate [23,24]. There should be a certain distance near 10 micrometers between the diffraction gratings at which the NSPP does not completely attenuate. The period of the diffraction gratings should be comparable with the carrier wavelength of the exciting pulse.

If a short electromagnetic pulse falls to the first diffraction grating, then the NSPP wave is generated in the waveguide and converted at the second grating into electromagnetic radiation which is directed to the photodiode [23,24]. Thus, it is possible to determine experimentally that the linear or nonlinear SPPs are excited by the shot electromagnetic pulse in the plasmon waveguide, when we read at the photodiode one pulse as the soliton, or as the dispersion-broadened

Gaussian pulse, or as the periodic signal. It should be noted, that the solitary shot electromagnetic pulse excites the nonlinear SPP wave in the plasmon waveguide. One can estimate the amplitude of NSPP electric vector by the relation $\chi E^2 = 4\pi\chi_3 E^2$, where $\chi_3 = 0.1 \times 10^{-15}\ m^2/V^2$ at $\lambda_0 = 800\ nm$ [20]. Normal component $E_x$ of the NSPP electric vector of pulse at the surface of plasmon waveguide at exciting by the short electromagnetic pulse with the intensity $\Phi \approx 0.5\ mJ/cm^2$, duration $30\ fs$ and center wavelength $\lambda_0 = 800\ nm$ ($\omega_0 = 2.36 \times 10^{15} s^{-1}$) at the medium parameters $\varepsilon_D = 1$, $\varepsilon_M = -26.82 + i1.69$, $\alpha_M = 4.15 \times 10^7\ m^{-1}$ has the magnitude $E_x \propto 2.5 \times 10^8\ V/m$.

## 5. CONCLUSION

In this work it is theoretically investigated the dependence of the self-modulated NSPP flow on the influence of the Kerr-type nonlinearity with saturation. At the surface of a planar plasmonic waveguide, which is the heterostructure of nonmagnetic metallic and dielectric layers, the TM mode of the NSPPs is formed when the waveguide is excited by short power electromagnetic pulse. Under the influence of the Kerr nonlinearity of both media, the NSPP wave becomes self-modulated, and the shape of the envelope and the modulation period of the surface plasmon wave are determined by the intensity of the exciting electromagnetic pulse.

The single pulse excites the NSPP wave with the envelope in the form of either cnoidal or solitary wave. The solitary wave is the sum of light and dark solitons depending on the ratio of the parameters of the exciting electromagnetic field and the parameters of the media. The NSPP envelope wave profile and period are transformed with varying of the intensity of source and/or the conditions of NSPPs excitation.

## APPENDIX A

We assume that the components of the NSPP field with the plane wavefront at the interface between the conducting and dielectric homogeneous media do not change along the transverse $y$-axis, that is $\partial/\partial y \to 0$. From the Maxwell's equations $\nabla \times \boldsymbol{B} = c^{-1}\partial(\varepsilon \boldsymbol{E})/\partial t$ and $\nabla \times \boldsymbol{E} = -c^{-1}\partial \boldsymbol{B}/\partial t$, we obtain a system of equations for the TM wave components $E_x, E_z, B_y \sim \exp(\alpha x)$ of NSPP

$$\frac{1}{c}\frac{\partial(\varepsilon E_x)}{\partial t} + \frac{\partial B_y}{\partial z} = 0,\ \frac{1}{c}\frac{\partial(\varepsilon E_z)}{\partial t} - \alpha B_y = 0,\ \frac{\partial E_x}{\partial z} - \alpha E_z + \frac{1}{c}\frac{\partial B_y}{\partial t} = 0. \quad (A.1)$$

By excluding $B_y$ in the system of equations (A.1), we obtain

$$\frac{1}{c^2}\frac{\partial^2}{\partial t^2}(\varepsilon E_z) + \alpha \frac{\partial E_x}{\partial z} - \alpha^2 E_z = 0,\ \frac{\partial}{\partial z}(\varepsilon E_z) + \alpha \varepsilon E_x = 0, \quad (A.2)$$

where $\varepsilon(t,z) = \varepsilon_L + \chi(E_x^* E_x + E_z^* E_z)$, $\varepsilon_L = const$ is the dielectric constant of the medium, $\chi = (4\pi\chi_D, 4\pi\chi_M)$ is the Kerr susceptibility of the medium, $\alpha = (-\alpha_D, \alpha_M) = const$ is the transverse damping coefficient of NSPP in the dielectric (D) and in the metal (M).

In the linear regime at $\chi E^2 \to 0$, the monochromatic continuous signal $E_{ax}\exp(-i\omega t)$ and $E_{az}\exp(-i\omega t)$, (where $E_{ax} = const$ and $E_{az} = const$), excites a harmonic SPP wave with the components of the electric vector $E_x = E_{ax}\exp(-i\omega t + i\beta z)$ and $E_z = E_{az}\exp(-i\omega t + i\beta z)$. We obtain from the equation system (A.1) the equation $\alpha_L^2 = (Re\beta)^2 - \varepsilon_L \omega^2/c^2$, where $\beta = (\omega/c)[\varepsilon_D\varepsilon_M/(\varepsilon_D + \varepsilon_M)]^{1/2}$ is the propagation constant at the boundary condition $-\alpha_D\varepsilon_M = \alpha_M\varepsilon_D$ for linear SPP TM mode, $L=D,M$. In this case, by substituting the expressions for the components of the electric vector in the form $E_x = E_{ax}\exp[-i\omega(t - z/v)]$ and $E_z = E_{az}\exp[-i\omega(t - z/v)]$ into the system of equations (A.1), we find the velocity of the SPP harmonic wave $v = c/n_L = const$, where $n_L = [\varepsilon_L + \alpha_L^2 c^2/\omega^2]^{1/2}$.

In the moving frame of reference $\tau = t - z/v$, $v = const$, the system of equations (A.2) for the components of the electric vector $E_x$ and $E_z$ takes the form

$$\frac{1}{c^2}\frac{d^2}{d\tau^2}(\varepsilon E_z) - \frac{\alpha}{v}\frac{dE_x}{d\tau} - \alpha^2 E_z = 0,\ \frac{d}{d\tau}(\varepsilon E_z) - \alpha v \varepsilon E_x = 0. \quad (A.3)$$

Combining equations (A.3), we represent the system of equations in the form

$$\frac{dE_x}{d\tau} + \frac{1}{\varepsilon-n^2}\frac{d\varepsilon}{d\tau}E_x - \frac{\alpha c n}{\varepsilon-n^2}E_z = 0,\ \frac{dE_z}{d\tau} + \frac{1}{\varepsilon}\frac{d\varepsilon}{d\tau}E_z - \frac{\alpha c}{n}E_x = 0, \quad (A.4)$$

where $n^2 = c^2/v^2$, $\varepsilon(\tau) = \varepsilon_L + \chi(E_x^* E_x + E_z^* E_z)$.

The velocity of moving frame must be chosen as $v = c/n = const$. In the general case $n \neq n_L$ and, in the particular case when $n = n_L$, the expression in the factor denominators of the first equation of the system (A.4) takes the form $\Delta\varepsilon = \varepsilon - n_L^2 = \chi(E_x^* E_x + E_z^* E_z) - \alpha_L^2 c^2/\omega^2$, where $\omega = \omega_0$ is the pulse carrier frequency.

## APPENDIX B

The Eq. (7) for the module of NSPP electrical vector does not depend on the cyclic coordinate (phase) $\varphi$,

$$\left(\frac{dE}{d\tau}\right)^2 = 2\left[\tilde{E} + \frac{\Omega^2}{2\chi}ln(\Delta\varepsilon + \chi E^2)\right] - \frac{1}{E^2}\tilde{P}^2. \quad (B.1)$$

We find the module $E$ of NSPP electric vector by integrating Eq. (B.1) and inverting the integral

$$\tau = \frac{1}{2}\int\left\{2E^2\left[\tilde{E} + \frac{\Omega^2}{2\chi}ln(\Delta\varepsilon + \chi E^2)\right] - \tilde{P}^2\right\}^{-1/2} dE^2. \quad (B.2)$$

The phase $\varphi$ of NSPP electric vector is determined by Eq. (8)

$$\frac{d\varphi}{d\tau} = \frac{1}{E^2}\tilde{P}^2, \quad (B.3)$$

from which we obtain the expression

$$\varphi = \varphi_0 + \tilde{P}^2 \int_0^\tau E^{-2} d\tau \quad (B.4)$$

where we should substitute the module $E(\tau)$.

We find the integral in Eq. (B.2) for the module $E$ taking into account that $\chi E^2 \ll 1$. Expand the logarithm in the denominator of the integral in the Taylor series $ln(\Delta\varepsilon + \chi E^2) \approx ln(\Delta\varepsilon) + (\chi/\Delta\varepsilon)E^2 - (\chi^2/2(\Delta\varepsilon)^2)E^4$, where $\Delta\varepsilon = \varepsilon_L - n^2$, and substitute the effective potential $U = -\frac{\Omega^2}{2\chi}ln(\Delta\varepsilon) - \frac{\Omega^2}{2\Delta\varepsilon}E^2 + \chi\frac{\Omega^2}{4(\Delta\varepsilon)^2}E^4$ into the elliptic integral at Eq. (B.2). Then we obtain

$$\frac{1}{2}\int\left\{\left[2\tilde{E} + \frac{\Omega^2}{\chi}ln(\Delta\varepsilon)\right]E^2 + \frac{\Omega^2}{\Delta\varepsilon}E^4 - \chi\frac{\Omega^2}{2(\Delta\varepsilon)^2}E^6 - \tilde{P}^2\right\}^{-1/2} dE^2.$$

We represent this integral in the form

$$\frac{a^2}{\sqrt{2\chi}\Omega}\int\{b_1 E^2 + b_2 E^4 - E^6 - b_3\}^{-1/2} dE^2,$$

where $b_1 = \left[2\tilde{E} + \frac{\Omega^2}{\chi}ln(\Delta\varepsilon)\right]\frac{2(\Delta\varepsilon)^2}{\chi\Omega^2}$, $b_2 = \frac{2\Delta\varepsilon}{\chi}$, $b_3 = \frac{2(\Delta\varepsilon)^2}{\chi\Omega^2}\tilde{P}^2$, and rewrite the elliptic integral as $\int_x^{a_3}\frac{dx}{[(a_1-x)(a_2-x)(a_3-x)]^{1/2}}$, where $x = E^2$, $a_1 a_2 + a_1 a_3 + a_2 a_3 = -b_1$, $a_1 + a_2 + a_3 = b_2$, $a_1 a_2 a_3 = -b_3$, and $a_1 > a_2 > a_3 > x$. Then from the Eq. (B.2), we find $\frac{2}{\sqrt{a_1-a_3}}\left[F(\tilde{\varphi},\tilde{k})\right] = -\frac{\sqrt{2\chi}\Omega}{\varepsilon_L - n^2}\tau$, where $\tilde{\varphi} = \arcsin\sqrt{\frac{a_3-x}{a_2-x}}$, $\tilde{k} = \sqrt{\frac{a_1-a_2}{a_1-a_3}}$ is the modulus of elliptic integral of the first kind $F(\tilde{\varphi},\tilde{k})$ [30]. Inverting the elliptic integral $F(\tilde{\varphi}, \tilde{k})$, we find the square of NSPP electric vector at the interface of dielectric and metal

$$E^2 = \frac{a_3 - a_2 \text{sn}^2(\bar{\tau}, \tilde{k})}{\text{cn}^2(\bar{\tau}, \tilde{k})},$$

where $\bar{\tau} = \frac{\sqrt{\chi(a_1 - a_3)}}{\sqrt{2}\Delta\varepsilon}\Omega\tau$.

For $a_1 > a_2 = a_3 > x$, the integral under consideration is $\int_x^{a_2} \frac{dx}{(a_1 - x)^{1/2}(a_2 - x)}$. Then from the Eq. (B.2), we obtain $\frac{2}{\sqrt{a_2 - a_1}} arctan\left(\sqrt{\frac{a_1 - x}{a_2 - a_1}}\right) = \frac{\sqrt{2\chi}\Omega}{\Delta\varepsilon}\tau$, whence we find

$$E^2 = a_1 sch^2(\bar{\tau}) + a_2 tanh^2(\bar{\tau}),$$

where $a_1 = b_2 - 2a_2$, $a_2 = \left(b_2 + \sqrt{3b_1 + b_2^2}\right)/3$.

**Acknowledgements.** The author is grateful for support to the Russian Science Foundation (RSF) (19-72-20154), and T. Nurieva for help in the work.